\documentclass[a4paper]{svmult}
\usepackage{times}
\usepackage[dvips]{graphicx,epsfig}
\usepackage{amsmath,amsfonts,amssymb}
\usepackage{cite,url}
\usepackage{dcolumn}
\usepackage{color} 

\begin{document}

\frontmatter


\mainmatter

\title*{\boldmath Full Configuration Study of 
Light No-Core Nuclei with Novel Realistic $NN$ Interaction JISP16}
\author{\underline{A. M. Shirokov}\inst{1,2}
\and{J. P. Vary}\inst{2}
\and{P. Maris}\inst{2} }  

\authorrunning{A. M. Shirokov, J. P. Vary, and P. Maris}
\institute{Skobeltsyn Institute of Nuclear Physics, Moscow State University, Moscow, 119991 Russia
\and {Department of Physics and Astronomy, Iowa State University, Ames,
Iowa 50011, USA}}

\titlerunning{Full Configuration Study of 
Light No-Core Nuclei with JISP16 $NN$ Interaction}

\maketitle
\begin{abstract}
We introduce the no-core full configuration (NCFC) approach and present
 results for $^4$He,  $^{12}$C and $^{14}$F with the realistic $NN$
 interaction, JISP16.   We obtain ground state energies and their
 uncertainties through exponential extrapolations that we demonstrate
 are reliable in $^4$He where fully converged results are obtained.  We
 find $^{12}$C is overbound by $1.7$~MeV and we predict the
 yet-to-be-measured binding energy of $^{14}$F to be $70.2\pm 3.5$~MeV.
 The extrapolated spectrum of $^{14}$F is in reasonable agreement with
 known features of the $^{14}$B spectrum.

\end{abstract}
%

\section{Introduction and Motivation}
The rapid development of {\it ab-initio} methods for solving finite
nuclei has opened the range of nuclear phenomena that can be evaluated
to high precision using realistic nucleon-nucleon ($NN$) and three-nucleon
($NNN$) interactions, even interactions tied to QCD \cite{ORK94,N3LO}
where renormalization is necessary \cite{Navratil07}.   Here we present
methods for the direct solution of the nuclear many-body problem by
diagonalization in a sufficiently large basis space that converged
binding energies are accessed --- either directly or by simple
extrapolation.  We do not invoke renormalization. We choose a harmonic
oscillator (HO) basis with two basis parameters, the HO energy
$\hbar\Omega$ and the many-body basis space cutoff $N_{\max}$, defined
below.  We assess convergence in this 
$2D$ parameter space.  

Such a direct approach may be referred to as a ``No-Core Full
Configuration'' (NCFC) method.  Given the rapid advances in numerical
algorithms and computers, as well as the development of realistic
non-local $NN$ interactions \cite{Shirokov07} that facilitate
convergence, we are able to 
achieve converged results, either directly or through
extrapolation. That is, we do not need to soften the $NN$ interaction by
treating it with an effective interaction formalism.  Renormalization
formalisms necessarily generate many-body interactions that
significantly complicate the theory and are often truncated for that
reason.  Renormalization without retaining the effective many-body
potentials also abandons the variational upper bound characteristic that
we prefer to retain. We use a realistic JISP16 $NN$ interaction
\cite{Shirokov07,JISP16} designed to describe light nuclei without $NNN$
interactions. 


We adopt an $m$-scheme basis approach where the many-body basis states
are limited by the imposed symmetries --- parity and total angular
momentum projection $M$, as well as by the cutoff in the total
oscillator quanta above the minimum for that nucleus ($N_{\max}$).  In
natural parity cases, $M = 0$ (or $\frac12$) enables the simultaneous calculation of the entire spectrum for that parity and $N_{\max}$.  In many light nuclei, we obtain results for the first few increments of $N_{\max}$ and extrapolate calculated observables from a sequence of results obtained with these $N_{\max}$ values.

In our NCFC approach, 
the input Hamiltonian is independent of $N_{\max}$; the computer
requirements for a NCFC
calculation are the same as that of the {\it
ab-initio} No Core Shell Model (NCSM) \cite{NCSMC12} with a 2-body
Hamiltonian renormalized to the chosen $N_{\max}$ basis. The NCFC results
are obtained by taking the limit of $N_{\max}\to\infty$. 
Both the NCFC and NCSM approaches guarantee that all observables are
obtained free of contamination from spurious center-of-mass (CM) motion
effects.

\section{Hamiltonian, 
basis selection and method of solution}

In order to carry out the NCFC calculations, we require a realistic $NN$
interaction that is sufficiently weak at high momentum transfers that we
can obtain a reasonable convergence trend.  The conventional
Lee--Suzuki--Okamoto renormalization procedure of the {\it ab-initio}
NCSM \cite{NCSMC12} develops effective interactions that provide answers
close to experimental observations.  However, the convergence trend of
the effective interaction sequences with increasing $N_{\max}$ is not
uniform and leads to challenges for extrapolation.  Therefore, we select
the realistic $NN$ interaction, JISP16,  that produces spectra and other
observables in light nuclei that are in reasonable accord with
experiment \cite{Shirokov07}. We include the Coulomb interaction between
the protons. 

We cast the 
many-body problem with the ``bare'' interaction in the same
HO basis and with the same definition of the
cutoff as the {\it ab-initio} NCSM \cite{NCSMC12}. That is, the 
many-body finite matrix problem is defined by $N_{\max}$, the maximum number of
oscillator quanta shared by all nucleons above the lowest HO
configuration for the chosen nucleus.   The exact NCFC answer emerges in the
limit $N_{\max}\rightarrow \infty $. This definition of the cutoff
allows us to retain the same treatment of the 
CM constraint that eliminates spurious CM excitations as in the {\it
ab-initio} NCSM.  The Hamiltonian
matrix also depends on the HO energy, $\hbar\Omega$.  

Our approach satisfies the variational principle and guarantees
uniform convergence from above the exact eigenenergy with increasing
$N_{\max}$. That is, the 
results for the energy of lowest state of each
spin and parity, at any $N_{\max}$ truncation, are upper bounds on the
exact NCFC converged answers and the convergence is monotonic with increasing
$N_{\max}$.   Our goal is to achieve independence of the two parameters
as that is a signal for convergence --- the result that would be
obtained from solving the same problem in a complete basis.  

We employ the code ``Many Fermion Dynamics --- nuclear'' (MFDn)
\cite{Vary92_MFDn} that evaluates the many-body Hamiltonian and obtains
the low-lying eigenvalues and eigenvectors using the Lanczos algorithm.    

By investigating the calculated binding energies of many light nuclei as
a function of the two basis space parameters, we determined that, once
we exclude the  $N_{\max} = 0$ result, the calculated points represent
an exponential convergence pattern.  Therefore, we fit an exponential
plus constant to each set of results as a function of $N_{\max}$,
excluding $N_{\max} = 0$ at fixed $\hbar\Omega$, using the relation: 
\begin{equation}
E_{\rm gs}(N_{\max})=a \exp (-cN_{\max}) + E_{\rm gs}({\infty}).
\label{exp+const}
\end{equation}

\section{\boldmath Extrapolating the ground state energy --- NCFC test
 cases with $\rm ^4He$}

\begin{figure}[b!]
\centerline{\includegraphics[width=9cm]{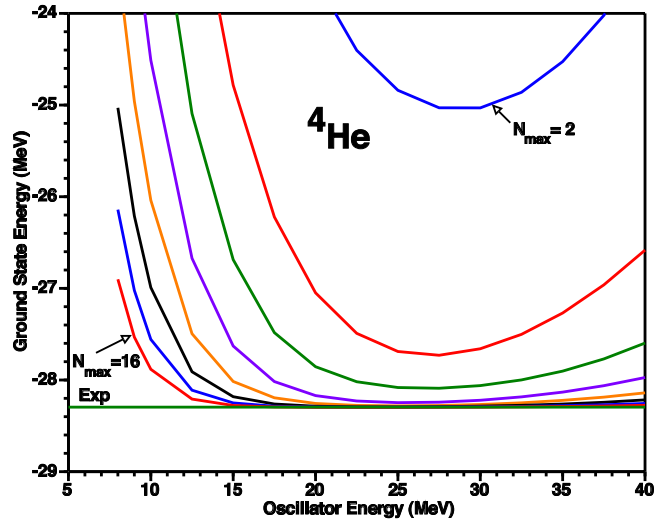}}
\caption{
Calculated ground state energy of $^4$He as
 function of the oscillator energy, $\hbar\Omega$,  for a sequence of
 $N_{\max}$ values. The curve closest to experiment corresponds to the
 value $N_{\max} = 16$ and successively higher curves are obtained with
 $N_{\max}$ decreased by 2 units for each curve.} 
\label{4He_gs_vs_hw}
\end{figure}

\begin{figure}
\centerline 
{\includegraphics[width=9cm]{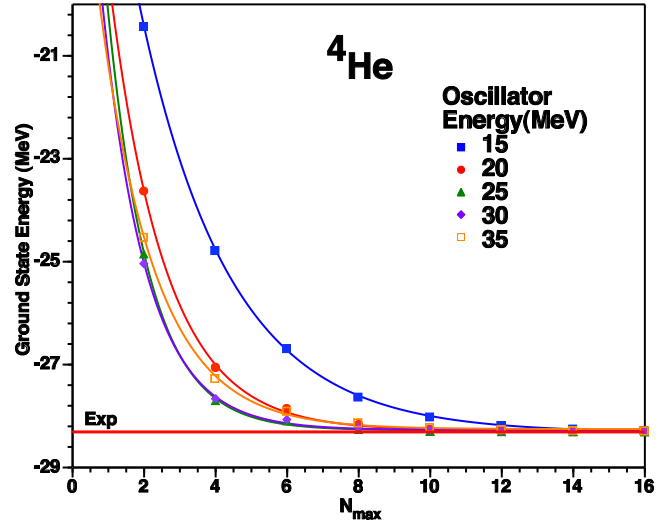}}
\caption{
Calculated ground state energy of $^4$He for
 $N_{\max}=2{-}16$ for JISP16 at selected values of $\hbar\Omega$. Each
 set of points at fixed $\hbar\Omega$ is fitted by Eq. (\ref{exp+const}) 
producing the solid curves.  Note the expanded
 energy scale.  Each point is a true upper bound to the exact
 answer. The asymptotes $ E_{\rm gs}({\infty})$ are the same to within 35 keV of their average
 value and they span the experimental ground state energy.} 
\label{4He_gs_range2_16}
\end{figure}

\begin{figure}[t]
\centerline 
{\includegraphics[width=9cm]{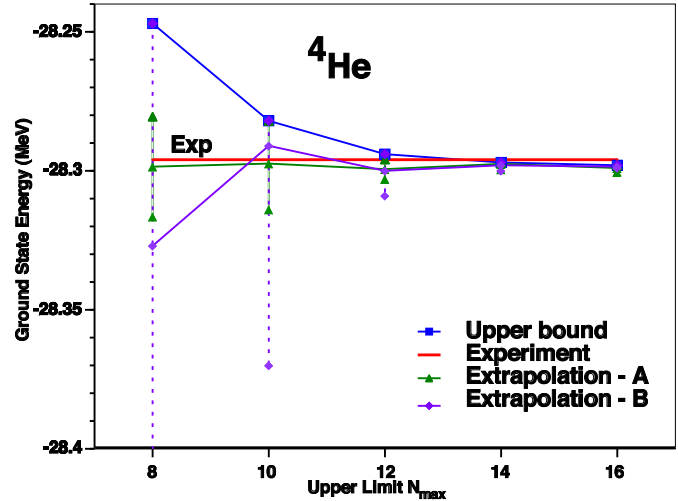}}
\caption{
Extracted asymptotes and upper bounds as 
 functions of the largest value of $N_{\max}$ in each set of points used
 in the extrapolation.  Four (three) successive points in $N_{\max}$ are
 used for the extrapolation A (B). Uncertainties are determined as described
 in the text.   Note the expanded scale and the consistency of the
 asymptotes as they fall well within their uncertainty ranges along the
 path of a converging sequence. } 
\label{4He_gs_extrap_insert_AB}
\end{figure}

We now investigate the convergence rate for the ground state energy as a
function of $N_{\max}$ and $\hbar\Omega$ for $^4$He where we also
achieve nearly exact results by direct diagonalization for comparison.
In particular, we present the results and extrapolation analyses for $^4$He in Figs. \ref{4He_gs_vs_hw} through \ref{4He_gs_extrap_insert_AB}.




The sequence of curves in Fig. \ref{4He_gs_vs_hw} for $^4$He illustrates
the trends we encounter in  calculations when evaluating the ground
state energy with the ``bare'' JISP16 interaction.  Our purpose with
$^4$He is only to illustrate convergence trends. The $N_{\max} = 18$
curve reaches to within 3 keV of the exact answer that agrees with
experiment.  

Next, we use these $^4$He results to test our ``extrapolation method A''
as illustrated in Fig. \ref{4He_gs_range2_16}.  For extrapolation A, we
will fit only four calculated points at each value of $\hbar\Omega$.
However, in Fig. \ref {4He_gs_range2_16} we demonstrate the exponential
behavior over the range $N_{\max} = 2 {-} 16$.  Later, we will introduce a
variant, ``extrapolation method B'' in which we use only three
successive points for the fit.  For extrapolation A, we select the
values of $\hbar\Omega$ to include in the analysis by first taking the
value at which the minimum (with respect to $\hbar\Omega$) occurs along
the highest $N_{\max}$ curve included in the fit, then taking one
$\hbar\Omega$ value lower by 5 MeV and three $\hbar\Omega$ values higher
by successive increments of 5 MeV.  For heavier systems we take this
increment to be 2.5 MeV. Since the minimum occurs along the
$N_{\max}=16$ curve at  $\hbar\Omega = 20$ MeV as shown in
Fig. \ref{4He_gs_vs_hw}, this produces the 5 curves spanning a range of
20 MeV in $\hbar\Omega$ shown in Fig. \ref{4He_gs_range2_16}.   

We recognize that this window of results in $\hbar\Omega$ values is
arbitrary.  Our only assurance is that it seems to provide a consistent
set of extrapolations in the nuclei examined up to the present time. 

For the resulting 5 cases shown in Fig. \ref{4He_gs_range2_16}, we
employ an independent exponential plus constant for each sequence,
perform a linear regression for each sequence at fixed
$\hbar\Omega$, and observe a small spread in the extrapolants that is
indicative of the uncertainty in this method.  Note that the results in
Fig. \ref{4He_gs_range2_16} are obtained with equal weights for each of
the points.  

For extrapolation A, we will fit sets of 4 successive points due to a desire to minimize the fluctuations
due to certain ``odd-even'' effects. 
These effects may be interpreted as sensitivity to incrementing the
basis space with a single HO state at a time while including two
successive basis states affords tradeoffs that yield a better balance in
the phasing with the exact solution.  

Next, we consider what weight to assign to each calculated point. We
argue that, as $N_{\max}$ increases, we are approaching the exact result
from above with increasing precision.  Hence, the importance of results
grows with increasing $N_{\max}$ and this should be reflected in the
weights assigned to the calculated points used in the fitting procedure.
With this in mind, we adopt the following strategy: define a chisquare
function to be minimized and assign a ``sigma'' to each calculated
result at $N_{\max}$ that is based on the change in the calculated
energy from the previous $N_{\max}$ value.  To complete these sigma
assignments, the sigma for the first point on the $N_{\max}$ curve is
assigned a value three times the sigma calculated for the second point
on the same fixed-$\hbar\Omega$ trajectory.  

As a final element to our extrapolation A strategy, we invoke the
minimization principle to argue that all curves of results at fixed
$\hbar\Omega$ will approach the same exact answer from above.  Thus all
curves will have a common asymptote and we use that condition as a
constraint on the chisquare minimization. 

When we use exponential fits constrained to have a common asymptote and
uncertainties based on the local slope, we obtain curves close to those
in Fig. \ref{4He_gs_range2_16}.  The differences are difficult to
perceive in a graph so we omit presenting a separate figure for them in
this case. It is noteworthy that the equal weighting of the linear
regression leads to a spread in the extrapolants that is modest.

The sequence of asymptotes for the $^4$He ground state energy, obtained
with extrapolation A, by using successive sets of 4 points in $N_{\max}$
and performing our constrained fits to each such set of 4 points, is
shown in Fig. \ref{4He_gs_extrap_insert_AB}.  We employ the independent
fits similar to those in Fig. \ref{4He_gs_range2_16} to define the
uncertainty in our asymptotes.  In particular, we define our
uncertainty, or estimate of the standard deviation for the constrained
asymptote,  as one-half the total spread in the asymptotes arising from
the independent fits with equal weights for each of the 4 points.  
In some other nuclei, on rare
occasions, we obtain an outlier when the linear regression produces a
residual less than 0.999 that we discard from the determination of the
total spread. Also, on rare occasions, the calculated upper uncertainty
reaches above the calculated upper bound.  When this happens, we reduce
the upper uncertainty to the upper bound as it is a strict limit. 

One may worry that the resulting extrapolation tool contains several
arbitrary aspects and we agree with that concern.  Our only recourse is
to cross-check these choices with solvable NCFC cases as we have done
\cite{Maris08}.  
We seek consistency of the constrained extrapolations as
gauged by the uncertainties estimated from the unconstrained
extrapolations described above.  Indeed, our results such as those shown
in Fig. \ref{4He_gs_extrap_insert_AB},  demonstrate that consistency.
The deviation of any specific constrained extrapolant from the result at
the highest upper limit $N_{\max}$ appears well characterized by the
assigned uncertainty.  We have carried out, and will present elsewhere, 
a far more extensive set of
tests of our extrapolation methods \cite{Maris08}.

As we proceed to applications in heavier nuclei, we face the technical
limitations of rapidly increasing basis space dimension.  In some cases,
only three points on the $N_{\max}$ curves may be available so we
introduce extrapolation B.  Our extrapolation B procedure uses three
successive points in $N_{\max}$ to determine the exponential plus
constant.  We search for the value of $\hbar\Omega$ where the
extrapolation is most stable and assign the uncertainty to be the
difference in the ground state energy of the highest two points in
$N_{\max}$.  As expected, since extrapolation B uses less ``data'' to
determine the asymptote, it will have the larger uncertainty.  Again, we
trim the upper uncertainty, when needed, to conform to the upper bound. 

We present the behavior of the asymptotes determined by extrapolations A
and B  in Fig. \ref{4He_gs_extrap_insert_AB} along with the experimental
and upper bound energies.  In this case the results are very rapidly
convergent at many values of $\hbar\Omega$ producing uncertainties that
drop precipitously with increasing $N_{\max}$ as seen in the figure. We
note that the uncertainties conservatively represent the spread in the
asymptotes since all the extracted asymptotes are consistent with each
other within the respective uncertainties. The largest $N_{\max}$ points
define the results quoted in Table \ref{tabA_2-16}, a ground state
overbound by $3 \pm 1$ keV.

\begin{table}\vspace{-1ex}
\caption{Binding energies 
of several light nuclei from experiment and theory. 
The theoretical results are obtained with the JISP16 interaction in 
NCFC calculations 
as described in the text.  The uncertainties in the rightmost digits of
 an extrapolation is quoted in parenthesis. 
We present also variational lower bounds for the 
binding energies and
the uppermost value of $N_{\max}$ used in the quoted extrapolations.}
\label{tabA_2-16}
{
\begin{tabular}{|c|c|c|c|c|c|}
\hline
Nucleus/property & Exp  & Extrap (A) &  Extrap (B) & Lower bound &Max$(N_{\max})$ \\ 
\hline

$^{3}$H~$|E \big(\frac 12^+, \frac 12 \big)|$ [MeV]   &   8.482
 & 8.369(1)   & 8.3695(25) & 8.367 & 18 \\
\hline

$^{3}$He~$|E \big(\frac 12^+, \frac12\big)|$ [MeV]   &   7.718           & 7.665(1)   & 7.668(5) & 7.663 & 18  \\
\hline

$^{4}$He~$|E (0^+,0)|$ [MeV]          & 28.296       & 28.299(1)   & 28.299(1) & 28.298 & 18   \\

\hline

$^{6}$He~$|E (0^+,1)|$ [MeV]          & 29.269       & 28.68(12)   & 28.69(5)  & 28.473 &14  \\

\hline

$^{6}$Li~$|E (1^+,0)|$ [MeV]          & 31.995       & 31.43(12)   & 31.45(5) & 31.185 &14   \\

\hline
$^{8}$He~$|E (0^+,2)|$ [MeV]          & 31.408        & 29.74(34)   & 30.05(60) & 28.927 & 12   \\
\hline
$^{12}$C~$|E (0^+_1,0)|$ [MeV]          & 92.162        & 93.9(1.1)  & 95.1(2.7) & 90.9 & 8  \\

\hline
$^{13}$O~$|E \big(\frac 32^-, \frac 32 \big)|$ [MeV]   & 75.556
 &75.6(2.2) &77.6(2.0) &69.1 & 8\\ 
\hline

$^{14}$B~$|E(2^-,2)|$ [MeV]          & 85.42 & 83.7(3.1) &85.5(2.0)
 &76.0 & 8\\
\hline
$^{14}$F~$|E(2^-,2)|$ [MeV]          & ?  
 &70.2(3.5) &71.8(2.4) &61.4 &8 \\
\hline

$^{16}$O~$|E (0^+_1,0)|$ [MeV]          & 127.619      & 143.5(1.0)       & 150 (14)  &134.5 & 8  \\
\hline

\end{tabular} }
\end{table}

\vspace{-1.5ex}

\section{\boldmath Extrapolating the ground state energy: 
NCFC for $\rm ^{12}C$ and $\rm ^{14}F$}\vspace{-.5ex}

In our investigations of the lightest nuclei \cite{Maris08} 
we observe a
marked correlation between binding energy and convergence rate: the more
deeply bound ground states exhibit greater independence of $\hbar\Omega$
at fixed $N_{\max}$.   Our physical intuition supports this correlation
since we know the asymptotic tails of the bound state wave functions
fall more slowly as one approaches a threshold for dissociation.  This
same intuition tells us to expect Coulomb barriers and angular momenta
to play significant roles in this correlation.

\begin{figure}
\centerline 
{\includegraphics[width=9cm]{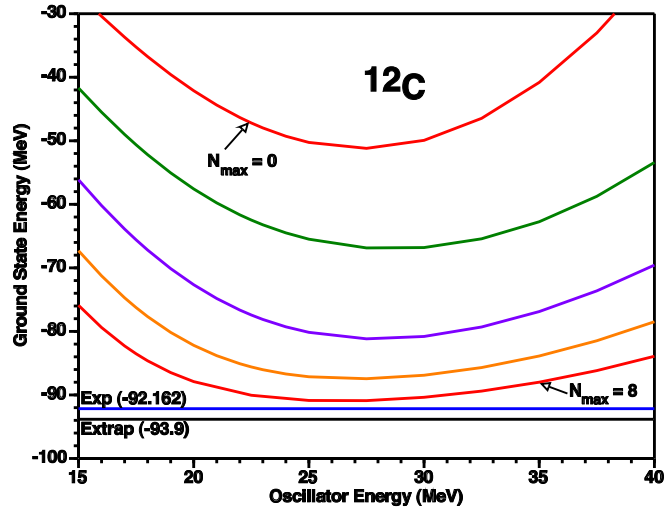}}
\caption{
Calculated ground state energy of $^{12}$C as function of the oscillator
 energy, $\hbar\Omega$, for selected values  of  $N_{\max}$. The curve
 closest to experiment corresponds to the value $N_{\max} = 8 $ and
 successively higher curves are obtained with $N_{\max}$ decreased by 2
 units for each curve.} 
\label{12C_gs_vs_hw}
\end{figure}

\begin{figure}
\centerline 
{\includegraphics[width=9cm]{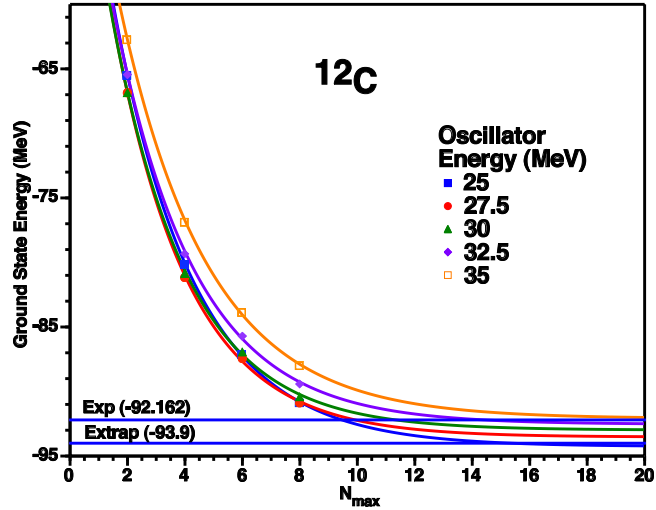}}
\caption{
Calculated ground state energy of $^{12}$C for
 $N_{\max}=2{-}8$ at selected values of $\hbar\Omega$ as described in
 the text. For each $\hbar\Omega$ the data are fit to an exponential
 plus a constant, the asymptote.   The figure displays the experimental
 ground state energy and the common asymptote obtained in extrapolation A.} 
\label{12C_gs_extrap}
\end{figure}

We proceed to discuss the  $^{12}$C results by introducing Figs. \ref{12C_gs_vs_hw} and \ref{12C_gs_extrap}.
The $^{12}$C nucleus is the first case for which we have only the
extrapolation from the ${N_{\max}=2{-}8}$ results since the  $N_{\max}=10$
basis space, with a dimension of 7,830,355,795, is beyond our present
capabilities.  Thus, in order to illustrate the details of our
uncertainties, we depict in Fig. \ref{12C_gs_extrap} the linear
regression analyses  of our results spanning the minimum in
$\hbar\Omega$ obtained at $N_{\max}=8$. Extrapolation A produces
overbinding by about 1.7 MeV.

\begin{figure}
\centerline{\includegraphics[width=9cm]{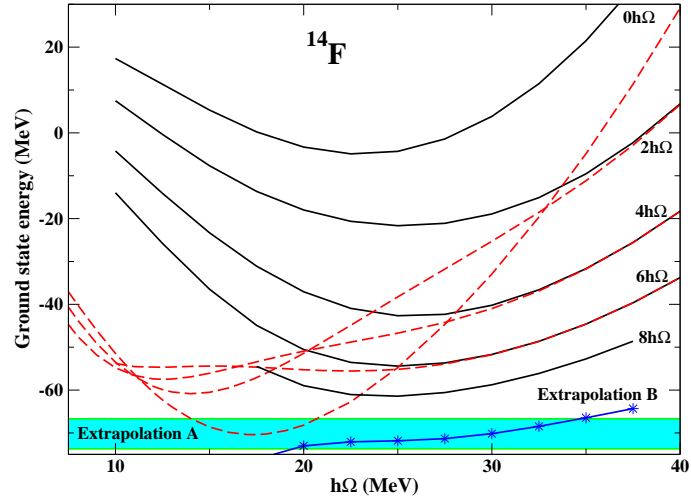}} 
\caption{
Calculated ground state energy of $^{14}$F for $N_{\max}=0{-}8$ with ``bare''
(solid lines) and effective (dashed lines) JISP16 interaction as
 function of the oscillator energy $\hbar\Omega$. Shaded area shows a
 confidence region of extrapolation A predictions, stars depict
 predictions by extrapolation B for individual $\hbar\Omega$ values.}
\label{14F_gs}
\end{figure}

Our next example is $^{14}$F, an exotic neutron-deficient nucleus, 
the first observation of which is
expected in an experiment planned in the Cyclotron Institute at Texas
A\&M University. In this case, we also attain the results up through
$N_{\max}=8$ presented in Fig. \ref{14F_gs}. 
The $N_{\max}=8$ basis
includes 1,990,061,078 states.
The results of
extrapolation B are shown in the figure by crosses for different
$\hbar\Omega$ values. In the case of extrapolation A, 
we obtain the binding
energy prediction of  $70.2\pm 3.5$~MeV
(shaded area in
Fig. \ref{14F_gs}) which is seen to be in a good correspondence with
extrapolation B that provides the binding energy of $71.8\pm 2.4$~MeV
with a smaller value of estimated uncertainty. It is interesting that,
contrary to our NCFC 
approach,  the trend of the conventional effective
interaction calculations of the binding energy is
misleading in this case: the minimum of the respective $\hbar\Omega$
dependence is seen from Fig. \ref{14F_gs} to shift up with increasing
$N_{\max}$ indicating the development of a shallow minimum at $N_{\max}=6$ around
$\hbar\Omega=12.5$~MeV; the ground state energy at this minimum  is
above the upper bound resulting from the variational principle and
calculations with the ``bare'' JISP16 interaction. The reliability of
the $^{14}$F results is supported by calculations of the
binding energy of the mirror nucleus  $^{14}$B (see Table~\ref{tabA_2-16}).

\begin{figure}
\centerline{\includegraphics[width=9.9cm]{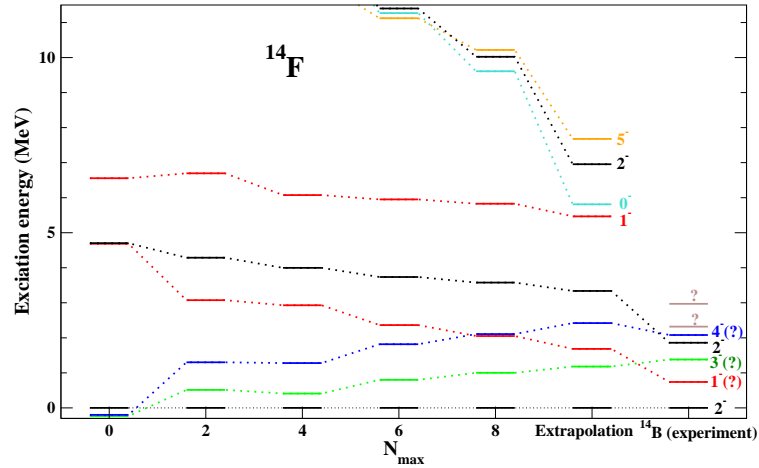}} 
\caption{
The $^{14}$F spectrum obtained with
 $N_{\max}=0{-}8$ and $\hbar\Omega=25$ MeV and by the extrapolation of
 the excited states in comparison with the spectrum of the mirror
 nucleus $^{14}$B.}
\label{14F_sp}
\end{figure}

We performed also calculations of the excited states in $^{14}$F. The
results obtained with $\hbar\Omega=25$ MeV in the range of  $N_{\max}$
values of 0{--}8, are presented in Fig. \ref{14F_sp}. We performed 
the extrapolation B for the energies of these states.  The respective
excitation energies, i.e. the differences between the
extrapolated energies and the extrapolated ground state energy, are also
shown in the figure. The  $^{14}$F spectrum is seen to be in a
reasonable agreement with the spectrum of the mirror nucleus
$^{14}$B. However we should note here that the spin assignments of nearly
all states in the $^{14}$B spectrum are doubtful.


\section{Conclusions and Outlook}\vspace{-1.5ex}

We present in Table \ref{tabA_2-16} a summary of the extrapolations
performed with methods introduced here and compare them with the
experimental results.  In all cases, we used the calculated results to
the maximum $N_{\max}$ available with the bare JISP16 interaction.   
Our overall conclusion is that these NCFC results demonstrate sufficient
convergence achieved for ground state energies of light nuclei 
allowing extrapolations to the infinite basis limit and estimations of their
uncertainties. These convergence properties are provided by the unique
features of the JISP16 $NN$ interaction.
The convergence rate reflects the short range
properties of the nuclear Hamiltonian.  Fortunately, new renormalization
schemes have been developed and applied that show promise for providing
suitable nuclear Hamiltonians based on other interactions with good convergence properties within
the NCFC method \cite{Bogner08}.  Additional work is needed to develop the
corresponding $NNN$ interactions. Also, further work is in progress  to
extrapolate the RMS radii.\\



We thank Richard Furnstahl, Petr Navr\'atil, Vladilen Goldberg,
Miles Aronnax and Christian Forssen for useful discussions. This work was
supported in part by the US Department of Energy Grants
DE-FC02-07ER41457 and DE-FG02-87ER40371. Results are obtained through
grants of supercomputer time at NERSC and at ORNL.   The ORNL resources
are obtained under the auspices of an INCITE award (David Dean, PI).  We
especially wish to acknowledge MFDn code improvements developed in
collaboration with Masha Sosonkina (Ames Laboratory), Hung Le  (Ames
Laboratory), Anurag Sharda (Iowa State University), Esmond Ng (LBNL),
Chao Yang (LBNL) and Philip Sternberg (LBNL).

\end{document}